\documentclass[pra, showkeys, showpacs, reprint]{revtex4-1} 

\usepackage{amsmath}  
\usepackage{amsfonts} 
\usepackage{graphicx} 

\begin{document}

\bibliographystyle{apsrev4-1}

\title{Experimental tests of Bertrand's question and the Duhem-Quine problem}
\author{Zhenning Liu}\email{zhenning.liu@durham.ac.uk}
\author{Charles S. Adams}\email{c.s.adams@durham.ac.uk}
\affiliation{Joint Quantum Center (JQC) Durham--Newcastle, Department of Physics,  Durham University, South Road, Durham, DH1 3LE, United Kingdom}
\date{\today}

\begin{abstract}
In this paper we report on an experimental test of Bertrand's question on the probability to find a random chord drawn inside a unit-radius circle with length greater than $\sqrt{3}$. In an experiment performed by tossing straws onto a circle, we confirm theoretical predictions that  the answer depends on the ratio of the circle diameter, $2R$, to the straw length, $L$, and that the special case corresponding to Laplace's principle of indifference is only obtained in the experimentally unattainable limit of infinite straw length, $\tilde{d}=2R/L\rightarrow 0$. In addition, we observe a systematic discrepancy in the limit, $\tilde{d}=2R/L\rightarrow 1$, where a large number of events are rejected. We conclude that the experimental test of Bertrand's paradox provides a good illustration of the Duhem-Quine problem---that hypothesis testing is always conditional on a bundle of real auxiliary assumptions. 
\end{abstract}


\maketitle

\section{Introduction} 

In 1889 Bertrand \cite{bert} asked the purely mathematical question: {\it If one draws a chord inside a circle at random, then what is the probability for its length to be longer than the side of the inscribed equilateral triangle?} He gave three answers, $1/4$, $1/3$ and $1/2$, depending on how we choose to interpret the words {\it at random}, see Fig.~1. In 1973 Jaynes \cite{jayn73} proposed a `solution' to this apparent `paradox' based on invariance to other assumptions---sometimes called `the principle of maximum ignorance'. Jaynes argued that the absence of additional information leads us to Laplace's {\it principle of indifference}---one of the `cornerstones' of probability theory, with wide-ranging significance across the physical sciences from statistical and quantum physics \cite{fuch13} to econophysics \cite{gari10}.  Note that in practice, there is often no reason to prefer an indifference probability distribution over any other \cite{wall96}.  
In reality, Jaynes asked a different question to Bertrand: {\it A long straw is tossed at random onto a circle; given that it falls so that it intersects the circle,
what is the probability that the chord thus defined is longer than a side of the inscribed equilateral
triangle?} This experimental question is distinct from Bertrand's purely mathematical question, \cite{mari94,arts14} and requires additional assumptions, such as, how `long' does the straw need to be relative to other length scales? The case of finite straw length has been analyzed theoretically by Porto {\it et al}. \cite{port11}, who showed that the correct answer depends on the ratio of the circle diameter, $2R$, to the straw length, $L$,  and only in the mathematical limit, $\tilde{d}=2R/L\rightarrow 0$, can we expect to recover Jaynes' `maximum ignorance' result.

\begin{figure}[t]
\begin{center}
\includegraphics[trim = 4mm 6mm 8mm 6mm, clip, width=8.5cm]{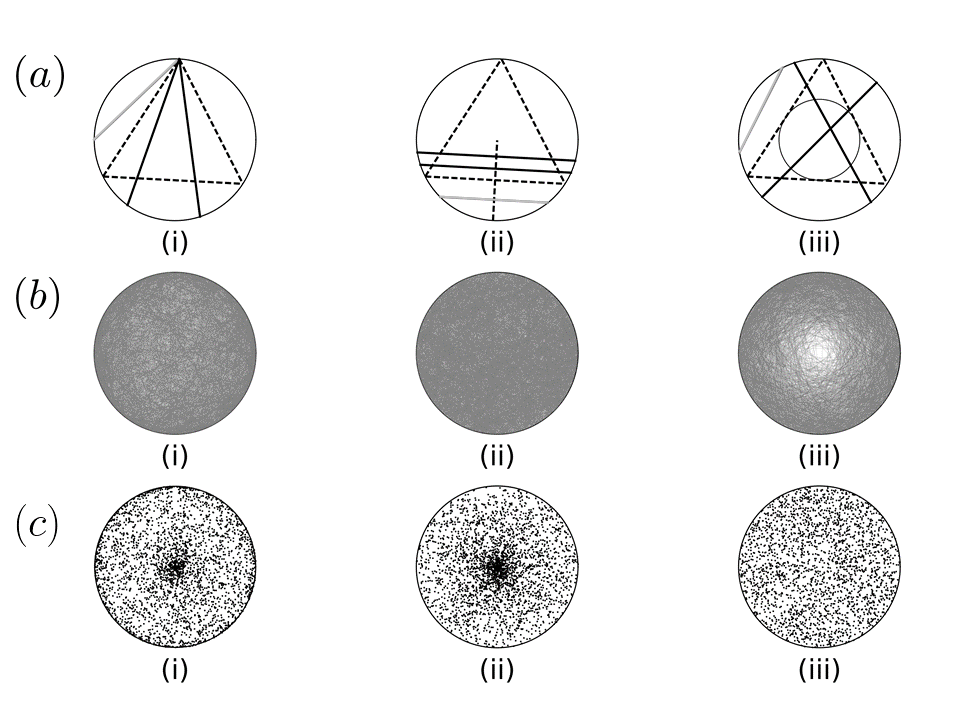}
\caption{Illustration of the three standard answers to Bertrand's question. (a) A chord drawn {\it at random} inside a unit-radius circle may be either longer (shown in black) or shorter (shown in grey) than the side length of an inscribed triangle, $\sqrt{3}$ (shown dashed). In case (i), the distribution of the chord end points is randomized giving a probability that a chord is longer than $\sqrt{3}$ equal to $1/3$. In case (ii), the radial position of the chord mid-point is randomized giving the answer $1/2$. In case (iii), both cartesian coordinates of the chord midpoint are randomized giving $1/4$. (b) and (c) show the corresponding chord and mid-points distributions, respectively, for each randomization procedure. In an experimental test with infinite straw length, there are no end points and no midpoint so only solution (ii) is allowed. 
}
\label{fig:1}
\end{center}
\end{figure}

The experimental form of Bertrand's question is a good example of the so-called Duhem Quine problem---the impossibility of testing a hypothesis without auxiliary assumptions \cite{duhem,stre01,stre06}. In practice, the design and analysis of an experiment requires many real assumptions.  In addition to the finite length of real straws, the experimental test of Bertrand's questions requires choices about how the random events are created, how to deal with failed attempts, and how we choose to record and analyze the data.  Each of these assumptions plays a role in determining the assumed distribution and hence the final answer. As Bertrand's `paradox' is both historically significant in the development of probability theory \cite{jayn73,mari94}, and more recently has become important in the context of an interesting class of metamaterials \cite{mara18}, such experimental and interpretative questions are significant. Although, Porto {\it et al}. \cite{port11} performed a numerical experiment, and there are related experimental studies on random coin tossing \cite{yong11}, there have been no rigorous experimental tests of Bertrand's question. As the experiment can be performed using simple equipment, exploit standard techniques in image and data analysis, and raises interesting interpretive question, it follows that systematic experimental investigations are long overdue.

In this letter, we perform an experimental test of Bertrand's question. We show how the practicalities of the experiment force us to address the additional assumptions in a way that may not arise in theory or simulation. For example, we observe a systematic discrepancy in the limit, $\tilde{d}=2R/L\rightarrow 1$, where a large number of events are rejected, which suggests that theoretically assumed distribution is not reproduced experimentally. We conclude that the experimental test of Bertrand's `paradox' is a useful test case to explore the Duhem-Quine problem that hypothesis testing is typically dependent on a bundle of interdependent assumptions.




\section{Experimental Test}






The principle of the experiment is illustrated in Fig.~2. We throw spinning straws of length $L$ onto a large sheet of card, Fig.~2(a), where a circle of radius $R$ is drawn, Fig.~2(b). Immediately obvious from Fig.~2(a) is that there are two distinct random aspects of straw throwing. First, the angle of the straw is randomized (isotropy or rotational randomness), and second the point where the straw lands (which we can take as its mid-point) is also random (spatial homogeneity or translational randomness). In an idealized scenario, we could imagine the experiment as an analogue simulator of translational and rotational randomness---randomly choosing a point where one end of the straw touches the floor, and second randomly choosing a direction in which the straw topples. In this respect there is a close analogy with Buffon's needle where a similar interplay between translational and rotational randomness is found \cite{port11}. 

To record each event, we take a photograph of the straw, Fig.~3(a), then draw a circle with a random position and diameter , Fig.~3(b). The experiment is repeated 3600 times and the resulting images form the complete dataset for analysis. The 3600 lines obtained from fitting the images are shown in Fig.~3(c). For any particular circle, we make a histogram of the distribution of chord lengths and calculate the probability that the chord length is longer than the side length of an inscribed equilateral triangle. We might expect that the experiment would produce one of the three standard answers to Bertrand's question shown in Fig.~1, and that the experiment will tell us which parameters are randomized in a random straw toss.
However, the answer is more complex and depends on how we choose to analyze the images. As both the circle size, $2R$, and position, can be varied in post processing this allows us to generate an infinite data set from the original finite set of images.  As highlighted by Porto {\it et al}. \cite{port11} a key parameter in the experiment is the circle diameter to length ratio, $\tilde{d}=2R/L$, and the answer to  Bertrand's question turns out to be a continuous function of this ratio \cite{mari94,port11}.

\begin{figure}[t]
\begin{center}
\includegraphics[trim = 0mm 0mm 0mm 0mm, clip, width=5.0cm]{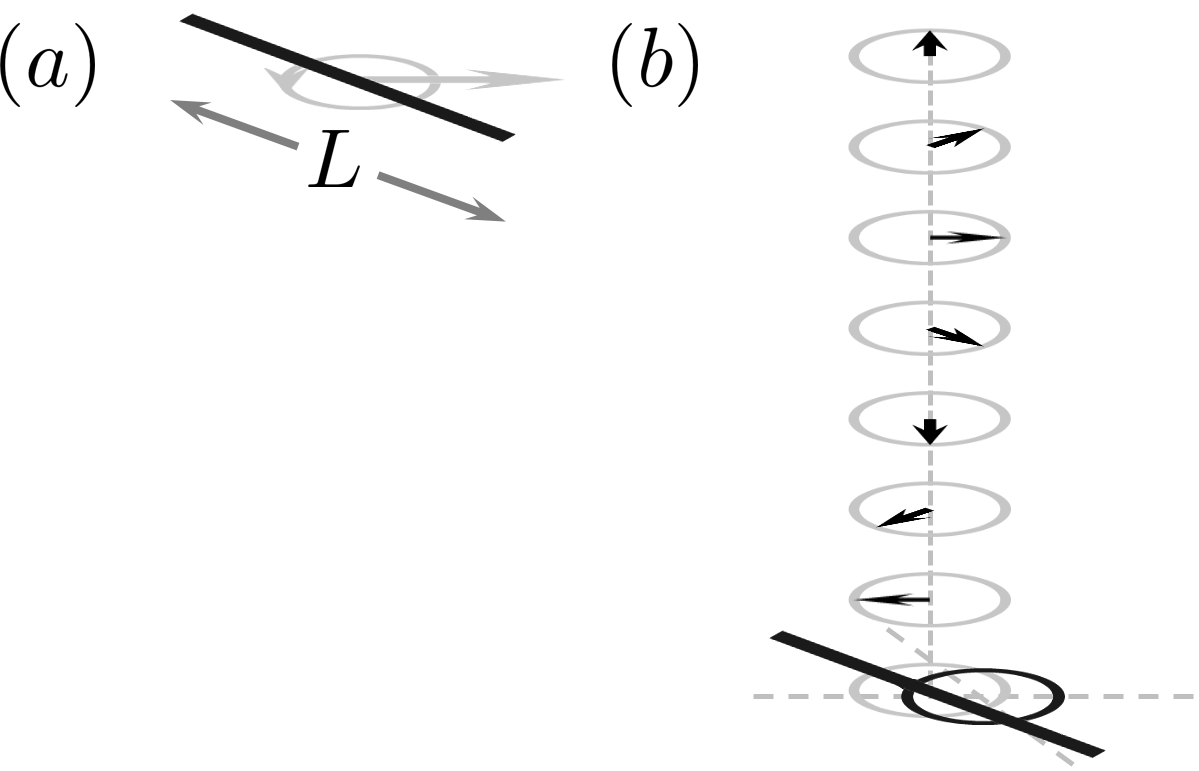}
\caption{Schematic of the experimental scheme for testing Bertrand's paradox. (a) A straw of length $L$ is thrown and spun such that both its position and angle on the circle are randomized. (b) An example showing the straw position after landing on a circle of radius $R$ drawn in black on a sheet of card.}
\end{center}
\label{fig:2}
\end{figure}

\begin{figure}[t]
\begin{center}
\includegraphics[trim = 3mm 4mm 4mm 4mm, clip, width=8.6cm]{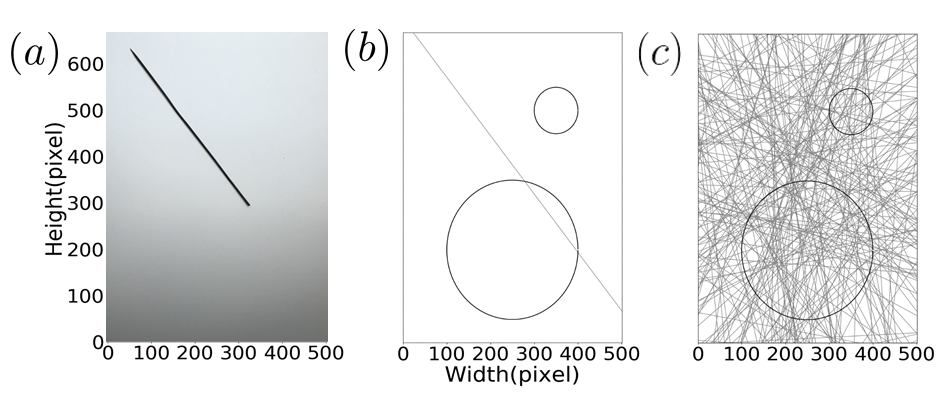}
\caption{(a) Camera image of the tossed straw on the paper. (b) The image processing routine fits a straight line through the path of the straw and adds a circle.
The position and size of the circle may be chosen in the analysis. In (b) we show two example circles to indicate, first a successful trial (the larger circle) and second an unsuccessful trial (smaller circle) where the straw misses the circle completely. Such cases are omitted from further analysis leading to a post-selection of `events'. The length of the line may either match the length of the straw or be extended as desired. (c) The experiment is repeated 3600 times and the path of each straw is added. In all plots, the coordinates on the axes are pixel counts of the camera. 
}
\end{center}
\label{fig:3}
\end{figure}

Not all trials produce a straw that completely overlaps with the circle as illustrated in Fig.~3(b), and it is necessary to post select events where a chord is defined. This post selection involves degrees of freedom about how we chose to analyze the data. 
In practice, we need to impose additional assumptions in order to obtain a result, and then we find that the result obtained depends on these assumptions. In selecting successful trials, there are three possible choices depending on whether we require zero, one or two intersections between the circle and the straw. We refer to these cases as the line, ray and segment viewpoints, respectively. The line viewpoint assumes that wherever the straw falls we can extend its length and if it crosses the circle, it is a valid trial. The ray viewpoint assumes that a straw that crosses the circle once can be extended to cross a second time. The segment viewpoint assumes that the trial is only valid if the straws makes two intersections with the circle.  The line, ray and segment viewpoints become equivalent in the mathematically ideal infinite-straw-length limit $\tilde{d}=2R/L\rightarrow 0$. For the data presented here, we use the segment viewpoint, however, it is possible to show the main qualitative conclusions are not dependent on this choice.
As in the experiment the length of straws is fixed, we modify the $R/L$-ratio (the dimensionless parameter $\tilde{d}$) by changing the radius of the circles in the analysis.
In practice, for each value of $R$, we draw 3000 randomly positioned circles with adjacent circles separated by a minimum of 20 pixels.
For an individual run, we measure the chord length, $\ell$, and bin the dimensionless length $x=\ell/(2R)$ in bins with width 0.02. The resulting normalized histograms for two values of the parameter $\tilde{d}$ are shown in  in Fig.~4. 

\begin{figure}[t]
\begin{center}
\includegraphics[trim = 0mm 0mm 0mm 0mm, clip, width=8.6cm]{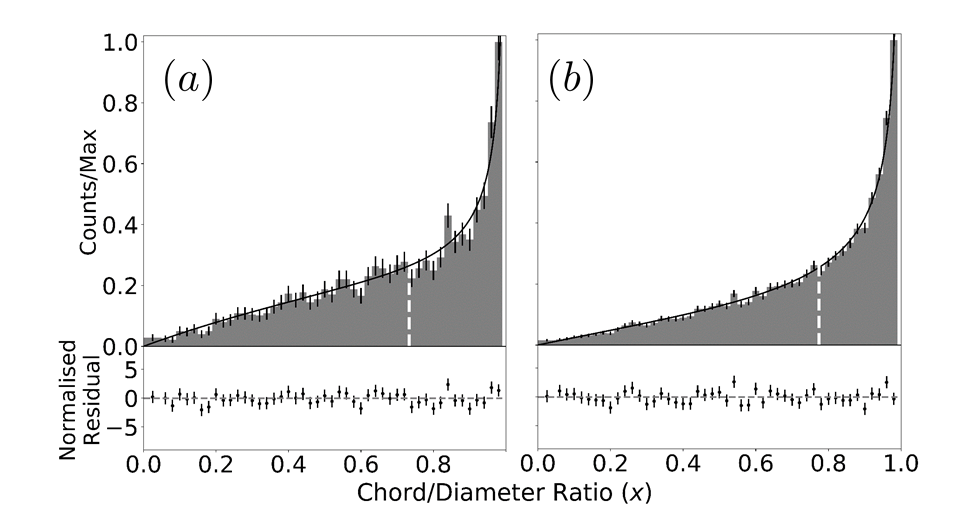}
\caption{The probability of measuring a chord length to diameter ratio between $x$ and $x+{\rm d}x$, for two values of the circle diameter to straw length ratio, $\tilde{d}=2R/L$: (a) For relatively short straws ($\tilde{d}=2R/L= 0.72$) and (b) for long straws ($\tilde{d}=2R/L= 0.112$). The geometrical solution, Eq.~(\ref{distri}) is shown as a solid line and the residuals are plotted below. The average chord length is indicated by the white dashed line.
}
\end{center}
\label{fig:4}
\end{figure}

\section{Theoretical Derivation}
The probability distributions shown in Fig.~4 follow from a simple geometrical argument.  Using rotational symmetry we can always rotate the chord axis to be vertical as in Fig.~5. We consider all possible positions of first touch that give a valid trial in the segment viewpoint. Using axial symmetry, we only need consider the left part of the first touch area. Let $p(x){\rm d}x$ be the probability to measure a chord with length between $2Rx$ and $2R(x+{\rm d}x)$, where $x$ is chord-to-diameter ratio, see Fig.~5(b). First we derive a probability cumulative function, $F(x)=\int_0^x p(x){\rm d}x$, by considering the area $A$ where all straws that first touch in $A$ generate a chord length shorter than $2Rx$. Using geometry, we obtain the following function for $A$:
\begin{equation}
A=LR\sqrt{1-x^2}-R^2\sin^{-1}(x)+R^2x\sqrt{1-x^2}~.
\end{equation}
The cumulative probability $F(x)=A/(A_1+A_2)$: 
\begin{equation}
F(x)=\frac{4\sqrt{1-x^2}-2\tilde{d}\sin^{-1}(x)-2x\tilde{d}\sqrt{1-x^2}}{4-\pi \tilde{d}}~.
\end{equation}
The probability density function $p(x)$ is given by the derivative of $F(x)$:
\begin{equation} \label{distri}
p(x){\rm d}x=\left(1- \frac{\pi \tilde{d}}{4}\right)^{-1}\left(\frac{x-x^2\tilde{d}}{\sqrt{1-x^2}}\right){\rm d}x~.
\end{equation}
This formula is used to plot the line in Fig.~4 and gives very good agreement with the data.
In the limit, $\tilde{d}\rightarrow 0$, one obtains Jaynes' distribution:\cite{jayn73}
\begin{equation}p(x){\rm d}x=\frac{x{\rm d}x}{\sqrt{1-x^2}}~.
\label{Jaynes}
\end{equation}
Eq. \eqref{Jaynes} is the limiting case of segment viewpoint but it is possible to show that the same distribution is obtained using the line or ray viewpoints.
The only difference is in how we determine the area of first touch. Instead in Fig.~5,  
for the ray viewpoint, the first-touch area is enclosed by upper half of the solid circle, the upper half of dotted circle and the two dotted lines. Repeating the analysis proves the equivalence between the ray and line viewpoints, and the limiting case of segment viewpoint in the limit $\tilde{d}\rightarrow 0$.

Comparing the experimental data of Fig.~4 to the predictions of Eq.~\eqref{distri}, we obtain a mean-square weighted deviation ($\chi^2_\mu$) \cite{hugh10} of $8.8$ and  $1.8$ for short and long straws, respectively. The experiment becomes a more accurate two-randomizer `machine' in the assumed limit of infinite straw length. The nature of the breakdown in the limit of large post-selection requires further investigation.  A formula for the average ratio of chord length to circle diameter, dashed lines in Fig.~4, using the segment viewpoint, can also be derived by integrating $xp(x)$ using \eqref{distri} from $0$ to $1$:
\begin{equation}
\overline{x}=\left(1-\frac{\pi }{4}\tilde{d}\right)^{-1}\left(\frac{\pi}{4}-\frac{2}{3}\tilde{d}\right)~.
\end{equation}
Note that in the limit $\tilde{d}\rightarrow 0$, the average chord length becomes $\overline{x}\rightarrow \pi/4$ .

From the distribution shown in Fig.~4, we can calculate the probability, $P$, that the chord length is longer that $\sqrt{3}R$ ($x>\sqrt{3}/2=0.866$). As apparent by comparing Fig.~4(a) and (b), $P$ depends on the circle diameter to straw length ratio $\tilde{d}=2R/L$. The measured probabilities for 18 different values of $\tilde{d}$ are plotted in Fig.~6. Again we can obtain an analytical result using geometry.
If we want the chord to be longer than $\sqrt{3}R$, then the position of first touch must be within the area $A_2$, and the probability of a chord length greater than $\sqrt{3}R$ is given by the ratio $A_2/(A_1+A_2)$, see Fig.~5(a). Using geometry
\begin{equation} \label{a1}
A_2=\frac{LR}{2}-\frac{\pi R^2}{6}-\frac{\sqrt{3} R^2}{4}~,
\end{equation}
\begin{equation} \label{a1a2}
A_1+A_2=LR-\frac{\pi R^2}{2}~,
\end{equation}
and the probability $A_2/(A_1+A_2)$ is
\begin{equation}
P(\tilde{d})=\frac{12-{2\pi \tilde{d}}-{3\sqrt3 \tilde{d}}}{24-6\pi \tilde{d}}~.
\label{eq:p_tilde_d}
\end{equation}
We plot this analytic expression together with the measurements in Fig.~6, and find reasonable agreement with the experimental data. However, there is a clear systematic discrepancy between theory and experiment in the limit $\tilde{d}\rightarrow 1$, where the amount of post-selection is maximal.
The reason for this discrepancy is unknown and will require further investigation. In the limit of infinitely long straws, $\tilde{d}\rightarrow 0$, we obtain  $P(\tilde{d})\rightarrow 0.5$ corresponding to Jaynes' result \cite{jayn73} as expected. 

\section{Discussion}

The main conclusion of Fig.~6 is that rather than one of these three standard answers to Bertrand's question (illustrated in Fig.~1), we find that the answer is a continuous function of the $\tilde{d}=2R/L$ ratio as predicted theoretically in Refs. \cite{mari94,port11}. In the limit of infinite straw length ($\tilde{d}=2R/L\rightarrow 0$) the answer tends towards 0.5.  In this limit the line, ray and segment viewpoints merge \cite{note}.  In the opposite limit $\tilde{d}\rightarrow 1$ s the amount of post-selection increases which reduces the probability of finding a chord longer than $\sqrt{3}R$.

\begin{figure}[t]
	\begin{center}
		\includegraphics[trim = 0mm 0mm 0mm 0mm, clip, width=8.6cm]{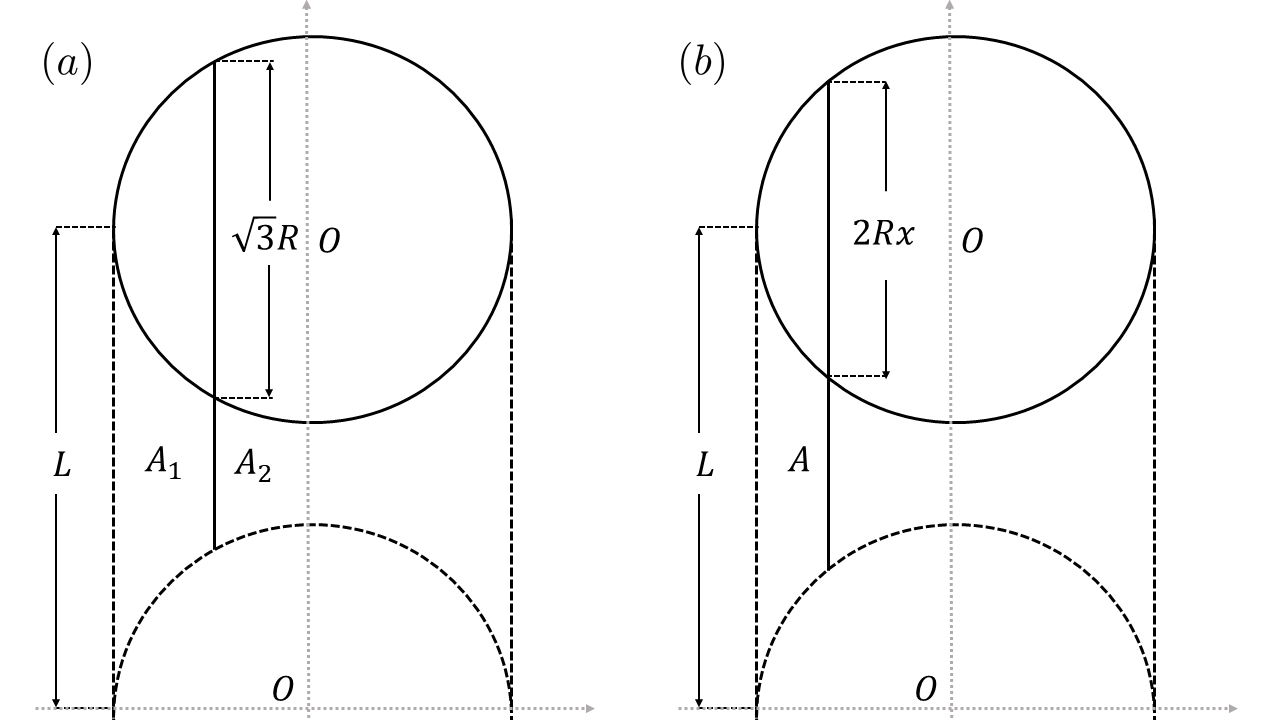}
		\caption{Geometry used to derive analytical expressions for the chord-length distribution. The vertical black line represents a straw with length $L$. The dashed circle trace out the far end-point for a valid trial in the segment viewpoint. (a) Sketch of chord with length equal to $\sqrt{3}R$. (b) Sketch of the general case with a chord of length $2Rx$. The probabilities of particular outcomes are related to the areas, $A_1$, $A_2$ and $A$.
		}
	\end{center}
	\label{fig:theo}
\end{figure}


\begin{figure}[t]
\begin{center}
\includegraphics[trim = 6mm 0mm 0mm 0mm, clip, width=7.0cm]{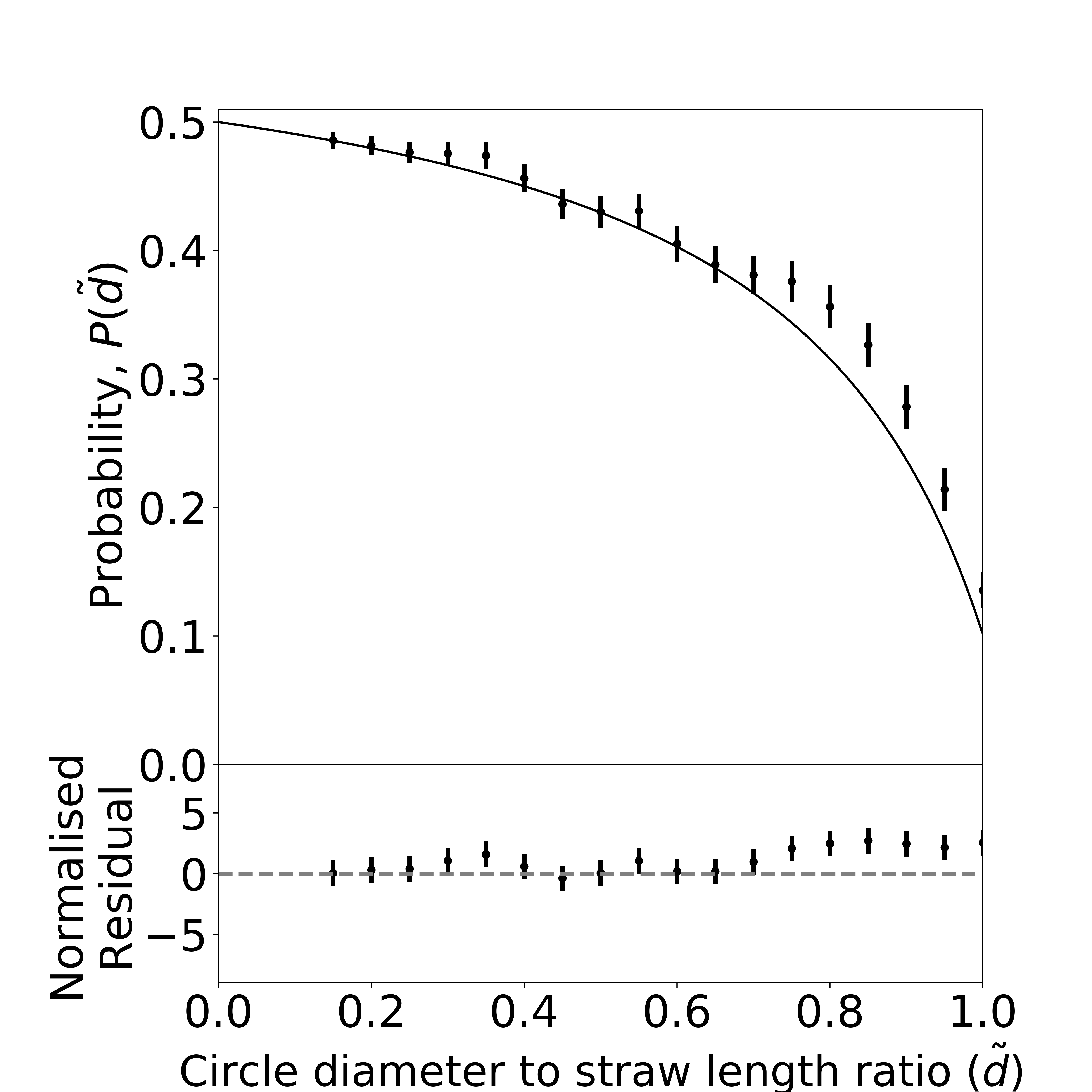}
\caption{The probability of a chord length longer than $\sqrt{3}R$ as function of the circle diameter to length ratio $\tilde{d}=2R/L$. The probability tends to 0.5 in the limit of infinite straw length or equivalently infinitely small circle diameter. The solid line is the theoretical prediction of Eq.~(\ref{eq:p_tilde_d}).
}
\end{center}
\label{fig:4}
\end{figure}

One interpretation of the Duhem-Quine thesis \cite{stre06} is that we cannot assess which assumptions, e.g. edge effects, that the experiment is a `perfect’ random number generator.  or that the straw length is sufficiently large,  is responsible for the discrepancy between the measurement and the expected result. In our example by simulating different straw lengths we are able to separate the effect of the straw length from other assumptions. The fact that we observe a systematic discrepancy in the limit, $\tilde{d}=2R/L\rightarrow 1$, suggest a residual bias in the experiment that cannot be eliminated simply by taking more data. The origin of this discrepancy is most likely due to the finite size of the raw images.


The experiment demonstrates the impracticality of testing Jaynes' maximal ignorance proposition.
In practice the experiment demands auxiliary conditions, i.e. a corresponding interpretative model, that are not independent from other assumptions about the relative straw length, image size, and post-selection.

\section{Summary}

In summary, we have performed an analysis of an experimental test of Bertrand's question that illustrates the fundamental nature of assumptions implicit in the scientific method, i.e. experimentation requires auxiliary conditions. Almost all experiments are designed with a prior probability distribution in mind. 
The experiment is in effect only a test of the assumed distribution, particularly in the case of post-selection. In the limit of long straw length the experimental test of Bertrand's paradox selects Laplace's principle of indifferent, like quantum experiments post-select quantumness, and Black-Scholes hedging selects gaussian fluctuations \cite{jame17}.  



\begin{acknowledgements}
We thank Nathan J. Cinnamond, Mia West, Robert Heighton for assistance with the experiment, in particular, obtaining the images used in the analysis, graphics and 
and for stimulating discussions. We also thank Tom Lancaster, Matthias C. M. Troffaes, Yuxun Zhang, Ifan Hughes and Robert Bettles for discussions, and Susan Hilton for technical support.
We acknowledge funding from Durham University. The data presented  in  this  paper  are  available  at \doi{10.15128/r2s1784k734}. 
\end{acknowledgements}


\begin{thebibliography}{58}%
\makeatletter
\providecommand \@ifxundefined [1]{%
 \@ifx{#1\undefined}
}%
\providecommand \@ifnum [1]{%
 \ifnum #1\expandafter \@firstoftwo
 \else \expandafter \@secondoftwo
 \fi
}%
\providecommand \@ifx [1]{%
 \ifx #1\expandafter \@firstoftwo
 \else \expandafter \@secondoftwo
 \fi
}%
\providecommand \natexlab [1]{#1}%
\providecommand \enquote  [1]{``#1''}%
\providecommand \bibnamefont  [1]{#1}%
\providecommand \bibfnamefont [1]{#1}%
\providecommand \citenamefont [1]{#1}%
\providecommand \href@noop [0]{\@secondoftwo}%
\providecommand \href [0]{\begingroup \@sanitize@url \@href}%
\providecommand \@href[1]{\@@startlink{#1}\@@href}%
\providecommand \@@href[1]{\endgroup#1\@@endlink}%
\providecommand \@sanitize@url [0]{\catcode `\\12\catcode `\$12\catcode
  `\&12\catcode `\#12\catcode `\^12\catcode `\_12\catcode `\%12\relax}%
\providecommand \@@startlink[1]{}%
\providecommand \@@endlink[0]{}%
\providecommand \url  [0]{\begingroup\@sanitize@url \@url }%
\providecommand \@url [1]{\endgroup\@href {#1}{\urlprefix }}%
\providecommand \urlprefix  [0]{URL }%
\providecommand \Eprint [0]{\href }%
\providecommand \doibase [0]{http://dx.doi.org/}%
\providecommand \selectlanguage [0]{\@gobble}%
\providecommand \bibinfo  [0]{\@secondoftwo}%
\providecommand \bibfield  [0]{\@secondoftwo}%
\providecommand \translation [1]{[#1]}%
\providecommand \BibitemOpen [0]{}%
\providecommand \bibitemStop [0]{}%
\providecommand \bibitemNoStop [0]{.\EOS\space}%
\providecommand \EOS [0]{\spacefactor3000\relax}%
\providecommand \BibitemShut  [1]{\csname bibitem#1\endcsname}%
\let\auto@bib@innerbib\@empty

\bibitem{bert} J. Bertrand, {\it Calcul des Probabilites}, Gauthier-Villars (Paris 1889).

\bibitem{jayn73} E. T. Jaynes, ``The Well-Posed Problem,'' Found. Phys. {\bf 3}, 477-493 (1973).

\bibitem{fuch13}
C. A. Fuchs and R. Schack, ``Quantum-Bayesian coherence,'' 
Rev. Mod. Phys. {\bf 85}, 1693-1715 (2013).

\bibitem{gari10} See e.g. Chapter 1 in U. Garibaldi and E. Scalas, {\it Finitary Probabilistic Methods in Econophysics}, CUP (Cambridge 2010).

\bibitem{wall96} P. Walley, ``Inferences from Multinomial Data: Learning about a bag of marbles (with discussion),'' J. Roy. Stat. Soc. Series B (Methodological), {\bf 58}, 3-57
(1996).


\bibitem{mari94}
L. Marinoff, ``A Resolution of Bertrand's Paradox,'' Philosophy of Science, {\bf 61}, 1-24 (1994).



\bibitem{arts14}
D. Arts and M. Sassoli de Bianchi, ``Solving the hard problem of Bertrand's paradox,'' J. Math. Phys. {\bf 55}, 083503 (2014).





\bibitem{port11}
P. Di Porto, B. Crosignani, A. Ciattoni and H. C. Liu, ``Bertrand's paradox: a physical way out along the lines of Buffon's needle throwing experiment,'' Eur. J. Phys. {\bf 32}, 819-825 (2011).

\bibitem{duhem}
P. Duhem,  {\it The Aim and Structure of Physical Theory}, Princeton University Press. 2nd. Ed., 1991.
W. V. O. Quine, ``Two Dogmas of Empiricism,'' The Philosophical Review. 60, 20-43 (1951).


\bibitem{stre01}
M. Strevens, ``The Bayesian Treatment of Auxiliary Hypotheses,'' Brit. J. Phil. of Sci. {\bf 52} 515-537 (2001). 


\bibitem{stre06}
M. Strevens, {\it Bayesian approach to philosophy of science}. In D.M.Borchert(ed.), Encyclopedia of Philosophy, second edition. Macmillan Reference USA, Detroit (2006). 



\bibitem{mara18} 
E. Marakis, M. C. Velsink, L. J. Corbijn van Willenswaard, R. Uppu, and P. W. H. Pinkse, ``Uniform line fillings'', arXiv:1809.01490.





\bibitem{yong11} E. H. Yong and L. Mahadevan, ``Probability, geometry, and dynamics in the toss of a thick coin,'' Am. J. Phys. {\bf 79}, 1195-1201 (2011).

\bibitem{hugh10} I. G. Hughes and T. P. A. Hase, {\it Measurements and their Uncertainties: A practical guide to modern error analysis}, OUP (Oxford 2010).


\bibitem{note}
Although Jaynes did not specify a viewpoint, in fact, he must have  assumed the line viewpoint in order to obtain Eq.~(\ref{Jaynes}) \cite{jayn73}.


\bibitem{jame17} J. James, {\it Quantitative Finance}
 IOP Publishing Ltd, Bristol (2017).



\end{thebibliography}

%

\end{document}